\documentclass[a4paper]{jpconf}
\usepackage{graphicx}
\usepackage{color}
\usepackage{amssymb}
\usepackage{amsfonts}

\begin{document}

\title{Pushing the Energy and Cosmic Frontiers with High-Energy Astrophysical Neutrinos\footnote{Based on an invited talk given at the 6th Symposium on Prospects in the Physics of Discrete Symmetries (DISCRETE 2018), Vienna, 26--30 November 2018}}

\author{Mauricio Bustamante}

\address{Niels Bohr International Academy \& DARK, Niels Bohr Institute, University of Copenhagen, Blegdamsvej 17, Copenhagen, Denmark}

\ead{mbustamante@nbi.ku.dk}

\begin{abstract}
The astrophysical neutrinos recently discovered by the IceCube neutrino telescope have the highest detected neutrino energies --- from TeV to PeV --- and travel the longest distances --- up to a few Gpc, the size of the observable Universe.  These features make them naturally attractive probes of fundamental particle-physics properties, possibly tiny in size, at energy scales unreachable by any other means.  The decades before the IceCube discovery saw many proposals of particle-physics studies in this direction.  Today, those proposals have become a reality, in spite of prevalent astrophysical unknowns.  We showcase examples of studying fundamental neutrino physics at these scales, including some of the most stringent tests of physics beyond the Standard Model. 
\end{abstract}


\section{Introduction}

High-energy astroparticle physics has two complementary facets.  One facet is oriented to astrophysics\ \cite{Ackermann:2019ows}.  Its goal is to find and study the most energetic astrophysical sources and discover the long-sought origin of ultra-high-energy cosmic rays (UHECRs), cosmic rays with energies in excess of $10^9$~GeV.  The other facet is oriented to particle physics\ \cite{Ackermann:2019cxh}.  Its goal is to probe the properties and interactions of fundamental particles at the highest energies, well beyond the reach of present-day particle accelerators.  Here we focus on the particle physics facet; in particular, on the unique strengths of high-energy astrophysical neutrinos to explore it.


\section{The fundamental-physics reach of high-energy astrophysical neutrinos}
\label{section:physics_reach}

Probes of fundamental physics that use high-energy particles of astrophysical origin crucially complement collider tests performed at lower energies.  In many proposed beyond-Standard-Model (BSM) theories, the size of the new effects grows with the energy of the interaction.  

The Large Hadron Collider probes physics at a center-of-momentum energy of $\sqrt{s} = 13$~TeV.  In comparison, UHECR interactions probe it at hundreds of TeV.  However, UHECR interactions on the cosmic microwave background make the Universe opaque to UHECRs, so the most energetic of them rarely reach Earth.  In addition, there are numerous unknowns related to UHECRs that complicate extracting fundamental physics from them, though it is still possible to do it\ \cite{AlvesBatista:2019tlv}.  High-energy astrophysical gamma rays are detected with energies of $\lesssim$100~TeV, and are believed to be produced with even higher energies.  However, because the Universe is also opaque to PeV photons, they are degraded to GeV--TeV upon reaching Earth.  

In contrast, high-energy astrophysical neutrinos --- so far detected in the TeV--PeV range --- reach Earth without being attenuated.  
There are at least three reasons why they are fitting probes of fundamental physics:
\begin{enumerate}
 \item
  {\bf They have the highest neutrino energies:}  The have energies that are orders of magnitude higher than accelerator neutrinos, which reach a few hundred GeV at most.  Thus, they can probe physics in the neutrino sector at $\sqrt{s} \sim {\rm TeV}$, comparable to Tevatron energies.  In the future, the detection of EeV neutrinos would allow to probe physics at $\sqrt{s} \sim 100~{\rm TeV}$, comparable to energies envisioned for the Future Circular Collider.
 \item
  {\bf They have the longest baselines:}  Even though the sources of high-energy astrophysical neutrinos remain undiscovered, the isotropy in their arrival directions hints at an origin in extragalactic sources located at Gpc-scale distances from Earth (1~Gpc $\approx$ $3 \cdot 10^{22}$~km), essentially the size of the observable Universe.  During the long trip, even tiny new-physics effects could compound to observable levels by the time the neutrinos reach Earth.
 \item
  {\bf They are weakly interacting:}  While the production mechanism and emission spectrum of neutrinos is unknown --- though there are solid theoretical expectations\ \cite{Anchordoqui:2013dnh, Ahlers:2018fkn} --- once neutrinos are emitted, they are not expected to interact during their propagation and until detected, due to their tiny cross sections.  However, new-physics effects could change this.  The absence of standard interaction effects during propagation removes one layer of uncertainty when looking for new physics. 
\end{enumerate}
There is one additional reason, arguably of practical importance.  A significant part of the science program of neutrino telescopes, like IceCube, is related to searching for the highest-energy, non-thermal astrophysical sources\ \cite{Ackermann:2019ows}.  The data that are recorded for that purpose can be directly repurposed for searching for fundamental physics\ \cite{Ackermann:2019cxh}.  

\begin{figure}[t!]
 \includegraphics[width=\textwidth]{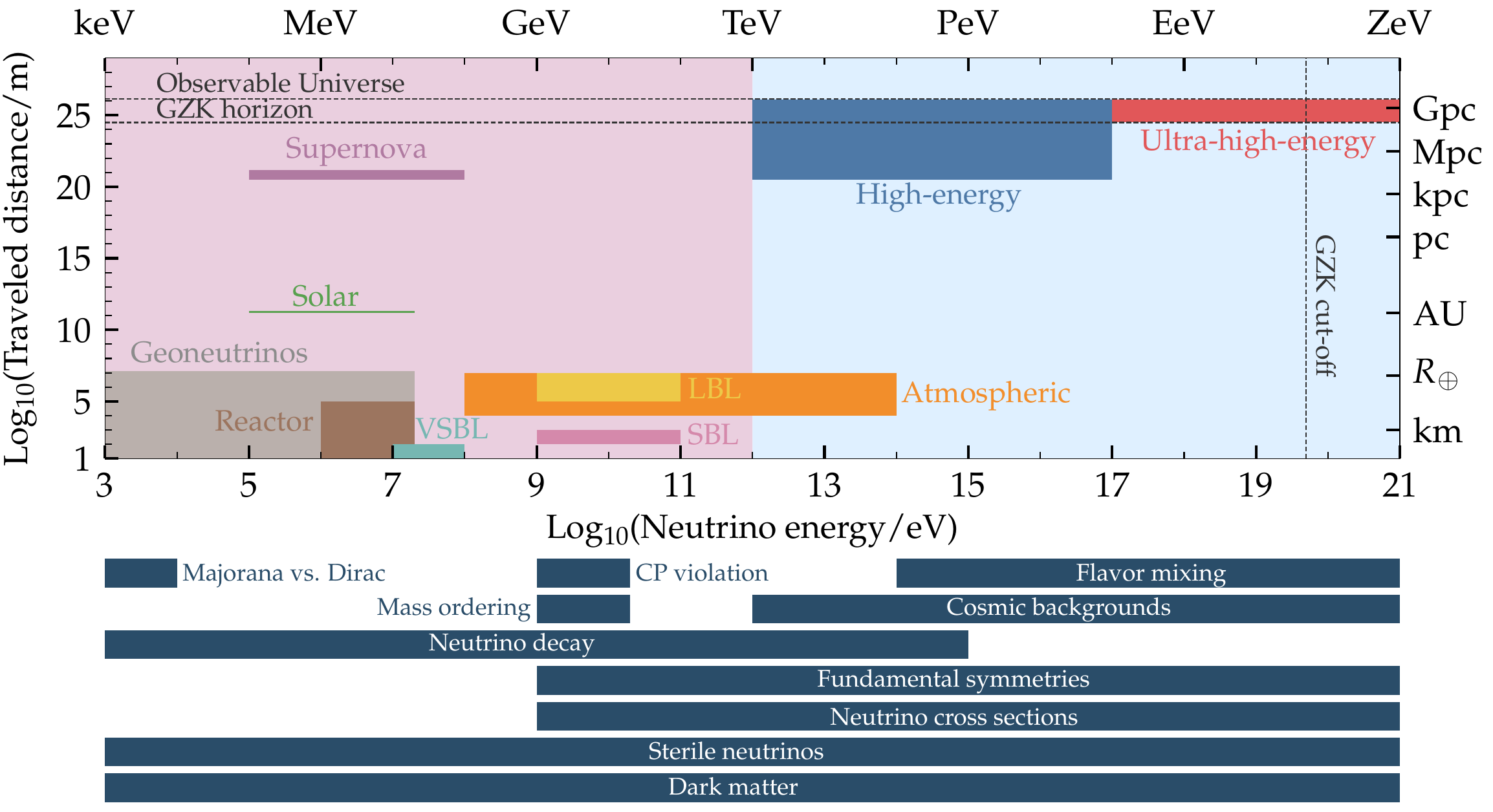}
 \caption{\label{label_scales}Neutrino energy and distance scales for different sources of neutrinos, and the physics models that they can test.  We focus on energies above TeV.  Figure reproduced from Ref.\ \cite{Ackermann:2019cxh}.}
\end{figure}

Figure\ \ref{label_scales} shows the distance and energy scales of neutrinos from different sources.  High-energy neutrinos reach the highest values on both axes.  The figure also shows a representative sample of BSM models that can be probed at high energies, complementing probes at lower energies.

Numerous new-physics effects grow as $\sim \kappa_n E_\nu^n L$, where $\kappa_n$ is a model-specific coupling strengths, $E_\nu$ is the neutrino energy, $L$ is the propagation distance, and $n$ is an integer that determines the energy-dependence of the effect.  For instance, for neutrino decays, $n = -1$, for CPT-odd violation of Lorentz-invariance, $n=0$, and for CPT-even violation of Lorentz invariance, $n > 1$.  Using astrophysical PeV neutrinos coming from sources located Gpc away, we can probe tiny couplings, {\it i.e.},
$\kappa_n \sim 4 \cdot 10^{-47} \left(E_\nu/{\rm PeV}\right)^{-n} \left(L/{\rm Gpc}\right)^{-1} {\rm PeV}^{1-n}$, an improvement of several orders of magnitude over limits obtained using atmospheric neutrinos\ \cite{Abbasi:2010kx, Abe:2014wla}.

\section{High-energy astrophysical neutrinos: basics and current status}
\label{section:background}

\subsection{Producing neutrinos}
\label{subsection:producing_nu}

Astrophysical neutrinos with TeV--PeV energies were discovered by IceCube only a few years ago\ \cite{Aartsen:2013bka, Aartsen:2013jdh}.  To date, their origin remains unknown.  We believe that they are made in cosmic accelerators that take protons and ions up to energies of EeV and above.  This may happen via collisionless shocks, wherein charged particles repeatedly cross shock fronts that move at relativistic speeds.  This produces an energy spectrum of protons (and ions) with an expected shape close to $\propto E_p^{-2}$, where $E_p$ is the proton energy.  

The high-energy protons interact with either photon fields\ \cite{Mucke:1999yb, Hummer:2010vx} ($p\gamma$) or surrounding matter\ \cite{Kelner:2006tc} ($pp$) in the sources\ \cite{Beresinsky:1969qj,Berezinsky:1975zz,Stecker:1978ah,Hill:1983xs,Yoshida:1993pt}.  To be concrete, below we consider $p\gamma$ interactions, in a simplified form.  When the proton energy and photon energy $E_\gamma$ satisfy the kinematic condition $E_p E_\gamma > 0.25$~GeV$^2$, their interaction produces a short-lived $\Delta^+(1232)$ resonance.  This happens, for instance, when protons of 0.1--100~PeV interact with MeV photons, which several classes of astrophysical sources have been seen to emit.  The $\Delta^+$ resonance decays into high-energy pions: $\Delta^+ \to p + \pi^0$, two-thirds of the time, and $\Delta^+ \to n + \pi^+$, one-third of the time.

The final-state neutrons escape the sources and decay into protons that are detected as UHECRs.  The final-state pions decay promptly: neutral pions make TeV--PeV gamma rays via $\pi^0 \to \gamma \gamma$, while charged pions make TeV--PeV neutrinos via $\pi^+ \to \mu^+ + \nu_\mu$, followed by $\mu^+ \to e^+ + \nu_e + \bar{\nu}_\mu$.  Gamma rays carry $10\%$ of the parent proton energy, and neutrinos carry $5\%$; thus, cosmic rays of tens of PeV make neutrinos and gamma rays of a few PeV.  The resulting neutrino energy spectrum is a power law whose slope, spectral breaks, and high-energy cut-off is inherited from the parent protons and photons\ \cite{Stecker:1978ah}.  Neutrino production in $pp$ interactions proceeds in a similar way, though in this case the neutrino spectrum is expected to be a power law with fewer features.  The fact that high-energy cosmic rays, gamma rays, and neutrinos are made in the same processes is at the core of multi-messenger astronomy. 

Unlike cosmic rays and gamma rays, neutrinos travel to Earth without losing energy to interactions with cosmic photon backgrounds.  Further, while cosmic rays are deflected by extragalactic and Galactic magnetic fields, neutrinos point back to their sources.  The bulk of the high-energy astrophysical neutrinos detected is compatible with coming from an isotropic distribution of extragalactic sources, of which several candidate source classes exist\ \cite{Anchordoqui:2013dnh, Ahlers:2018fkn}.  These candidates have been observed to emit gamma rays or are believed to harbor the necessary conditions to accelerate UHECRs. The contribution from neutrinos produced in sources inside the Milky Way, concentrated towards the Galactic Center, is of order $\sim$10\%, at most\ \cite{Ahlers:2015moa, Aartsen:2017ujz}.

\subsection{Detecting neutrinos}

The tiny cross sections that make neutrinos fitting probes of fundamental physics also make them challenging to detect.  IceCube monitors 1~km$^3$ of clear Antarctic ice using photomultipliers (PMTs) buried deep in it to search for the dim Cherenkov light that is created as a result of the interaction of a high-energy neutrino with the ice.  

When a TeV--PeV neutrino deep-inelastic-scatters off a nucleon in the ice, the interaction can be charged-current (CC), {\it i.e.}, $\nu_l + N \to l^- + X$, or neutral-current (NC), {\it i.e.}, $\nu_l + N \to \nu_l + X$, where $N$ is either a proton or neutron, $X$ are final-state hadrons, and $l = e, \mu, \tau$.  The final-state charged particles initiate showers that emit Cherenkov light that is recorded by the PMTs.  Anti-neutrinos undergo the same interactions, charge-conjugated. 

The CC interaction of a $\nu_e$ or $\nu_\tau$, and the NC interaction of a neutrino of any flavor, creates a ``cascade'' --- a particle shower whose light profile expands approximately radially out from the interaction vertex for several hundred meters.  Because the showers made by $\nu_e$ and $\nu_\tau$ look alike, it is difficult to tell these two flavors apart.  On the other hand, the CC interaction of a $\nu_\mu$ creates a ``track'' --- the final-state high-energy muon of the interaction propagates for up to several kilometers, leaving an elongated, clearly distinguishable track of Cherenkov light.

When looking for an astrophysical neutrino flux, the main background is from atmospheric neutrinos and muons made when cosmic rays interact in the upper atmosphere of Earth.  The energy spectrum of atmospheric neutrinos follows that of the cosmic rays: it is roughly $\propto E_\nu^{-3.7}$.   Beyond 100~TeV, the harder spectrum of astrophysical neutrinos becomes dominant.  

After about 7.5 years of running, IceCube has detected 103 showers and tracks that start inside the instrumented volume, 60 of which have energies above 60 TeV\ \cite{Taboada_Neutrino2018:Talk}.  These contained events constitute evidence of astrophysical neutrinos with a statistical significance of $> 7\sigma$.  A separate data set of about 1000 astrophysical through-going tracks, that are born outside the detector and cross part of it, provides independent evidence with a similar statistical significance\ \cite{Aartsen:2017mau}.  The energy spectrum of astrophysical neutrinos is either $\propto E_\nu^{-2.15}$\ \cite{Aartsen:2016xlq} or $\propto E_\nu^{-2.5}$\ \cite{Aartsen:2015knd}, depending on what data set is used.


\section{Neutrino observables to probe fundamental physics}
\label{section:observables}

New-physics effects may affect one or more of the following high-energy neutrino observables:
\begin{itemize}
 \item 
  {\bf Energy:}  The neutrino energy spectrum is expected to be a relatively featureless power law.  However, BSM effects can introduce spectral features that are not expected from astrophysical production mechanisms, {\it e.g.}, dips and peaks due to new resonant interactions between high-energy neutrinos and low-energy relic neutrinos\ \cite{Ioka:2014kca, Ng:2014pca,Shoemaker:2015qul}, or to interactions between neutrinos and dark matter\ \cite{Bhattacharya:2019ucd}.
 \item
  {\bf Arrival direction:}  The bulk of the diffuse neutrino flux is expected to come from isotropically distributed extragalactic sources, and so it is expected to be isotropic itself.  Thus, anisotropies in the distribution of arrival directions towards large accumulations of matter --- the Galactic Center or dwarf satellite galaxies of the Milky Way --- could be indicative of BSM neutrino-matter or neutrino-dark matter interactions.  
 \item
  {\bf Flavor composition:}  The decay of pions produces neutrinos with flavor ratios $(\nu_e:\nu_\mu:\nu_\tau) = (1:2:0)$.  En route to Earth, the standard expectation is that neutrino oscillations transform the flavor composition into nearly $(1:1:1)$\ \cite{Learned:1994wg}.  Currently, measurements of flavor composition at IceCube are compatible with this expectation\ \cite{Aartsen:2015ivb, Aartsen:2015knd, Bustamante:2019sdb}.  Yet, the measurements have large uncertainties, due to how challenging it is to measure flavor\ \cite{Aartsen:2015ivb, Aartsen:2015knd}.  Numerous BSM models acting during neutrino propagation may introduce large deviations in the flavor composition\ \cite{Beacom:2003nh, Pakvasa:2007dc, Bustamante:2010bf, Bustamante:2010nq, Mehta:2011qb, Bustamante:2015waa, Arguelles:2015dca, Shoemaker:2015qul, Gonzalez-Garcia:2016gpq, Rasmussen:2017ert, Ahlers:2018yom}.  Representative possibilities are Lorentz- and CPT-invariance violation, neutrino decay, and new long-range neutrino interactions with matter, which we expand upon below.
 \item
  {\bf Timing:}  Modifications of the energy-momentum relation introduced by Lorentz-invariance violation cause neutrinos and photons of different energies, produced simultaneously in transient astrophysical phenomena, to have different speeds and arrive on Earth at different times\ \cite{Longo:1987ub, Wang:2016lne, Diaz:2016xpw, Boran:2018ypz}.  Similar delays could be induced by BSM scattering of high-energy neutrinos off the low-energy relic neutrino background\ \cite{Murase:2019xqi}.
\end{itemize}

There are significant astrophysical and cosmological uncertainties associated to high-energy astrophysical neutrinos.  Regardless, it is possible, already today, to extract fundamental physics from the above neutrino observables.  We explore this below by way of a few examples.


\section{Current status, open questions}
\label{section:status_open}

\begin{figure}[t!]
 \begin{minipage}[t]{0.492\textwidth}  
  \centering
  \begin{minipage}[c][8cm][c]{\linewidth}
  \includegraphics[width=\linewidth]{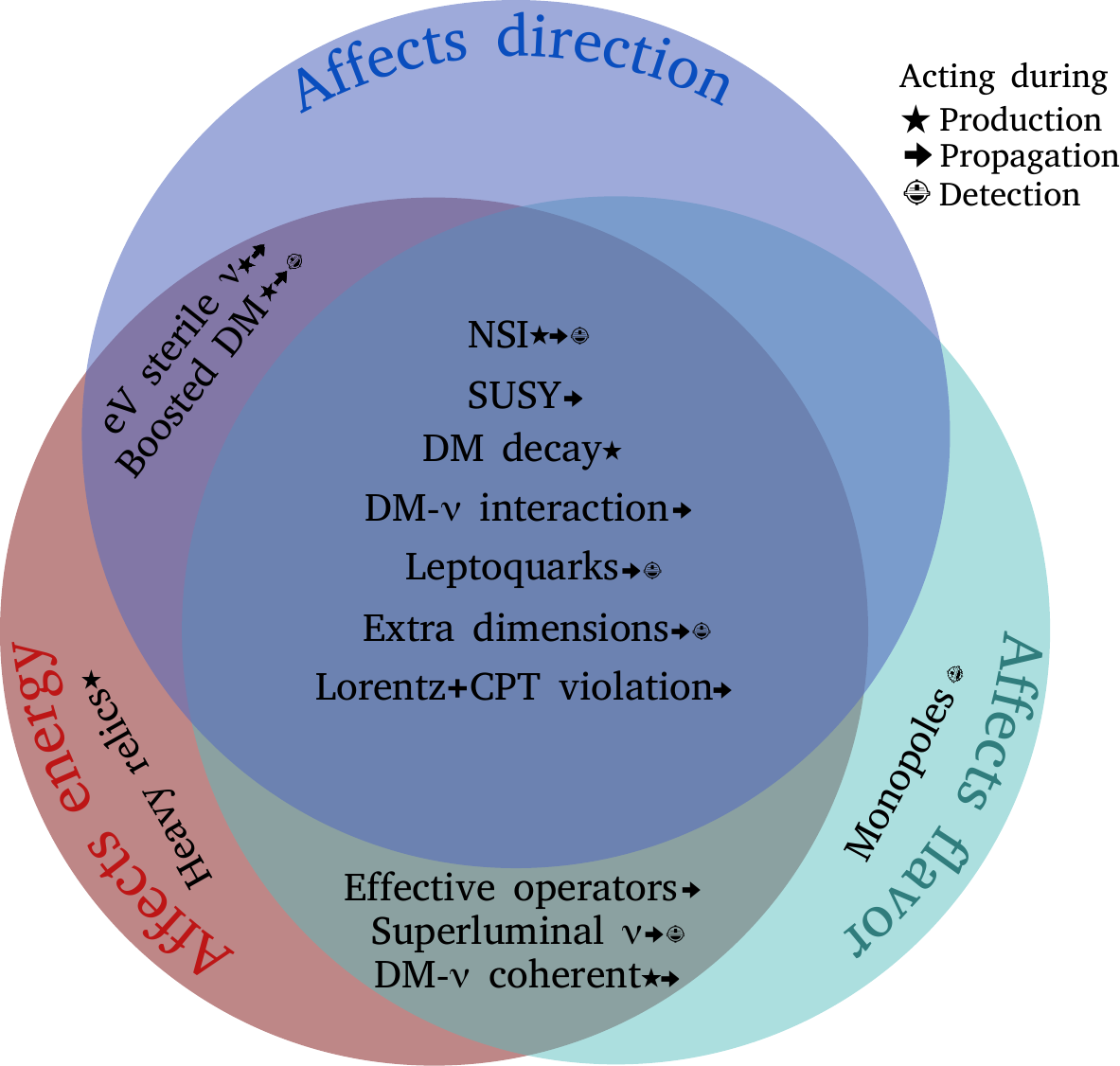}
  \end{minipage}
  \caption{\label{fig:classification}Non-exhaustive list of new-physics models classified by the observable that they affect and the stage where they act.  In this plot, we left out the timing observable.  Figure reproduced from Ref.\ \cite{Arguelles_review:in_prep}.}
 \end{minipage}
 \hfill
 \begin{minipage}[t]{0.492\textwidth}  
  \centering
  \begin{minipage}[c][8cm][c]{\linewidth}
  \includegraphics[width=\linewidth]{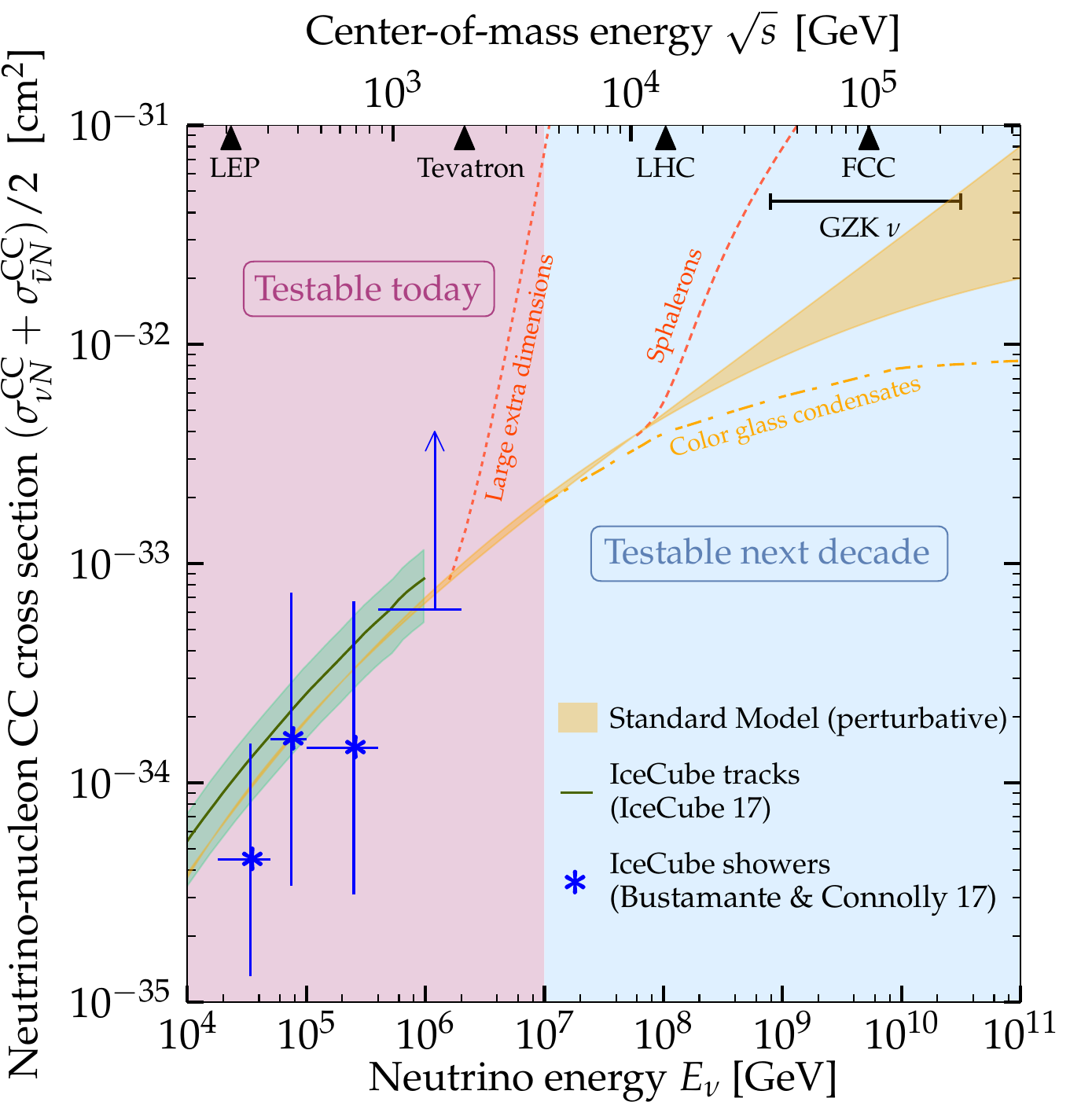}
  \end{minipage}
  \caption{\label{fig:cross_section}Measurements and predictions of the high-energy neutrino-nucleon cross section.  Figure reproduced from Ref.\ \cite{Ackermann:2019cxh}.}
 \end{minipage}
\end{figure}

Figure \ref{fig:classification} shows a possible classification of new-physics models that are testable using high-energy astrophysical neutrinos\ \cite{Arguelles_review:in_prep}.  Models are classified according to what neutrino observable from Section\ \ref{section:observables} they affect, and on whether they affect neutrinos during their production, propagation, or detection at Earth.  The list of models in Fig.\ \ref{fig:classification} is not exhaustive, but representative.

Below, we briefly expand on a few selected fundamental-physics measurements; see Ref.\ \cite{Ahlers:2018mkf} for a more complete presentation.

\begin{itemize}
 \item
  {\bf Neutrino-nucleon cross section:}  Until recently, the neutrino-nucleon cross section had been measured only up to neutrino energies of $\sim$350~GeV\ \cite{Tzanov:2005kr}.  The detection of high-energy neutrinos in IceCube has allowed us to measure the cross section at energies in the TeV--PeV range\ \cite{Aartsen:2017kpd, Bustamante:2017xuy, Aartsen:2018vez}.  Figure\ \ref{fig:cross_section} shows that the measurements are compatible with precise SM predictions\ \cite{CooperSarkar:2011pa}.  Yet, because the measurement uncertainties are still large, there is still room for small contributions from new physics.  At higher neutrino energies, still undetected, we could probe larger deviations from the SM predictions due to new physics --- {\it e.g.}, large extra dimensions\ \cite{AlvarezMuniz:2001mk} or electroweak sphalerons\ \cite{Ellis:2016dgb} --- and the internal structure of nucleons --- {\it e.g.}, color glass condensates\ \cite{Henley:2005ms}.
 \item
  {\bf Testing fundamental symmetries:}  The SM is an effective theory.  At energies above its energy scale, its symmetries may break down and new ones may appear.  Breaking CPT or Lorentz invariance\ \cite{Colladay:1998fq}, two of the linchpins of the SM, would affect multiple neutrino observables\ \cite{Kostelecky:2003cr, Hooper:2005jp, Kostelecky:2008ts, Bazo:2009en, Bustamante:2010nq, Kostelecky:2011gq, Gorham:2012qs, Borriello:2013ala, Stecker:2014xja, Tomar:2015fha, Amelino-Camelia:2015nqa, Liao:2017yuy}, increasing in strength with neutrino energy.  So far, tests of the violation of Lorentz invariance have been performed with intermediate-energy\ \cite{Abbasi:2010kx, Abe:2014wla} and high-energy\ \cite{Aartsen:2017ibm} atmospheric neutrinos.  Using astrophysical neutrinos of higher energies will increase the sensitivity further.
 \item
  {\bf Neutrino decay:}  Neutrino decay channels in the SM have associated lifetimes that are at least thirty orders of magnitude longer than the age of the Universe; for all practical purposes, in the SM neutrinos are stable\ \cite{Pal:1981rm, Hosotani:1981mq, Nieves:1983fk}.  However, BSM models may introduce new, faster decay modes; {\it e.g.}, via the emission of a Majoron\ \cite{Chikashige:1980qk, Gelmini:1982rr}.  Regardless of the specific BSM decay channel, the decay of the heavier neutrinos into the lighter ones, acting across cosmological distances, could leave sizable imprints in the high-energy neutrino energy spectrum and flavor composition\ \cite{Beacom:2002vi, Baerwald:2012kc, Shoemaker:2015qul, Bustamante:2016ciw, Denton:2018aml}.  The sensitivity of IceCube to these imprints yields neutrino lifetime constraints that are several orders of magnitude better than the ones obtained using neutrinos coming from closer sources\ \cite{Bustamante:2016ciw}.
 \item
  {\bf Dark matter:}  The decay or self-annihilation of dark matter into neutrinos\ \cite{Beacom:2006tt, Yuksel:2007ac, Feng:2010gw, Murase:2012xs}, or the interaction of high-energy neutrinos with dark matter in the Galactic Center\ \cite{Arguelles:2017atb}, could induce line-like features and low-energy pile-ups in the neutrino energy spectrum and an excess or deficit of events in the direction of large Galactic\ \cite{Adrian-Martinez:2015wey, Aartsen:2017ulx} and extragalactic\ \cite{Aartsen:2013dxa} concentrations of dark matter. The absence of these features in the data has placed significant constraints on the properties of dark matter, from light to superheavy\ \cite{Chianese:2017nwe, Bhattacharya:2019ucd}.
 \item
  {\bf Hidden neutrino interactions:}  BSM mediators may introduce new, ``secret'' interactions of high-energy astrophysical neutrinos with the relic neutrino background\ \cite{Ioka:2014kca, Ng:2014pca, Shoemaker:2015qul, Kelly:2018tyg}, with dark backgrounds\ \cite{Anchordoqui:2007iw, Klop:2017dim, Capozzi:2018bps}, or with cosmological distributions of ordinary matter\ \cite{Bustamante:2018mzu} that may significantly affect the neutrino energy spectrum and flavor composition.
 \item
  {\bf Sterile neutrinos:}  The mixing of high-energy active and sterile neutrinos during propagation over cosmological distances may induce deviations from the expectation of equi-flavor composition at Earth.  However, the stringent limits on active-sterile mixing angles from oscillation experiments constrain the deviations to be small, except in the unlikely scenario where the sources emit a significant amount of sterile neutrinos\ \cite{Brdar:2016thq}.
\end{itemize}


\section{Final remarks, and the future}
\label{section:conclusions}

In the TeV--PeV range, IceCube, its planned extension IceCube-Gen2\ \cite{Aartsen:2014njl}, and the planned KM3NeT\ \cite{Adrian-Martinez:2016fdl}  should provide a continuous stream of data in coming years.  The larger event statistics will improve measurements of neutrino cross sections and flavor composition.  This will be complemented by improvements in neutrino energy, direction, and flavor reconstruction.

In the EeV range, neutrinos are predicted\ \cite{Greisen:1966jv, Zatsepin:1966jv}, but remain undetected\ \cite{Zas:2017xdj, Aartsen:2018vtx, Gorham:2019guw}.  There is a guaranteed flux of ``cosmogenic'' EeV-scale neutrinos expected from the photohadronic interaction of UHECRs with the cosmic microwave background.  However, due to prevailing UHECR unknowns --- their mass composition and injection spectrum, and the redshift evolution of the number density of their sources --- the predictions of the cosmogenic neutrino flux shape and normalization are uncertain\ \cite{Kotera:2010yn, Ahlers:2012rz, Romero-Wolf:2017xqe, AlvesBatista:2018zui, Heinze:2019jou}.  Yet, if the EeV neutrino flux is high enough, a suite of envisioned next-generation ultra-high-energy neutrino telescopes\ \cite{AlvesBatista:2019tlv} could probe new physics at even higher energy scales and with tinier couplings, of $\kappa_n \sim 10^{-50}$~EeV$^{1-n}$.

Astroparticle physics lies at the root of the history of particle physics: the positron, muon, and pion were discovered in cosmic rays.  During the XX century, progress in particle physics was mainly driven by a progression of increasingly energetic and luminous particle colliders.  It is perhaps fitting that, at a time when colliders are at an impasse in finding how to extend the Standard Model, we seek again guidance in astroparticle physics.


\ack{MB is supported by the Villum Fonden project no.~13164.}


\section*{References}


\providecommand{\newblock}{}

\end{document}